\documentclass[aps,preprint,showpacs]{revtex4-1}
\usepackage{amsfonts}
\usepackage{amsmath}
\usepackage{amssymb}
\usepackage{graphicx}
\usepackage{rotating}
\usepackage{verbatim}
\usepackage{hyperref}
\usepackage{color}   
\usepackage[dvipsnames]{xcolor}
\usepackage{soul}

\begin{document}

\title{Magnetized color superconducting quark matter under compact star conditions: Phase structure within the SU(2)$_\mathbf{f}$ NJL model}

\author{M. Coppola$^{a,b}$, P. Allen$^{a}$, A.G. Grunfeld$^{a,b}$ and N.N. Scoccola$^{a,b,c}$}

\affiliation{$^{a}$ Department of Theoretical Physics, Comisi\'on Nacional de Energ\'ia
At\'omica, Av.Libertador 8250, 1429 Buenos Aires, Argentina~\\
 $^{b}$ CONICET, Rivadavia 1917, 1033 Buenos Aires, Argentina~\\
 $^{c}$ Universidad Favaloro, Sol\'is 453, 1078 Buenos Aires, Argentina}


\begin{abstract}
The properties of magnetized color superconducting cold dense quark matter under compact star conditions are investigated using an $SU(2)_f$ Nambu Jona-Lasinio (NJL)-type model in which the divergences are treated using a magnetic field independent regularization scheme in order to avoid unphysical oscillations. We study the phase diagram for several model parametrizations. The features of each phase are analyzed through the behavior of the chiral and superconducting condensates together with the different particle densities for increasing chemical potential or magnetic field. While confirming previous results derived for the zero magnetic field or isospin symmetric matter case, we show how the phases are modified in the presence of $\beta$-equilibrium as well as color and electric charge neutrality conditions.
\end{abstract}

\maketitle

\section{Introduction}

In the last decades, the QCD phase diagram as a function of the temperature and baryon chemical potential has been the focus of intense research~\cite{BraunMunzinger:2008tz,Fukushima:2010bq}. Particularly, during the last years many works in the literature were devoted to the study of quark matter under the influence of strong magnetic fields (see e.g.~\cite{Kharzeev:2012ph,Andersen:2014xxa,Miransky:2015ava} and refs. therein).
One of the reasons is that the estimated magnetic field created in relativistic heavy ion collisions is of the order $10^{18}-10^{20}$ G~\cite{Kharzeev:2007jp,Skokov:2009qp,Voronyuk:2011jd}. Another motivation is that, in the astrophysics scenario, certain compact objects called magnetars can have surface magnetic fields up to $10^{15}$ G~\cite{Duncan:1992hi,Paczynski:1992}, with estimates for the magnetic field values at their centers of a few orders of magnitude larger~\cite{Chatterjee:2014qsa,Lai:1991,Bandyopadhyay:1997kh,Ferrer:2010wz}. In that case, the relevant region of the QCD phase diagram is that of low temperature and intermediate values of density, where color superconducting phases of quark matter are expected to exist. At asymptotically large chemical potentials, the characteristics of the color superconducting phases can be analyzed using perturbative methods~\cite{Bailin:1984}. These methods, however, are not expected to be valid for the range of moderate densities relevant for magnetars. In addition, the well-known sign problem prevents the lattice QCD approach from being applied to this sector of the phase diagram~\cite{Karsch:2003jg,Ding:2015ona}.
In this situation, effective models of QCD arise as a powerful tool to circumvent these problems. One of the most popular effective models that preserves QCD chiral symmetries is the Nambu Jona-Lasinio (NJL) model~\cite{reports}. In this model, gluon degrees of freedom are integrated out in favor of some local quark-antiquark interactions and chiral symmetry is dynamically broken. When effective quark-quark interactions are added, quarks can form Cooper pairs that give rise to a variety of color superconducting phases~\cite{Buballa:2003qv,Alford:2007xm}.
In this context, the effect of a constant magnetic field has been analyzed by several authors~\cite{Ferrer:2005vd,Fukushima:2007fc,Noronha:2007wg,Fayazbakhsh:2010gc,Fayazbakhsh:2010bh,Mandal:2012fq,Mandal:2016dzg,Mandal:2017ihr}.
At this point, it is important to remark that the local character of the interactions considered in the NJL-type models leads to divergences in the momentum integrals. These divergences need to be handled in some way in order to completely define the model and yield meaningful results. Several regularization procedures are possible even in the absence of magnetic fields~\cite{reports}. Moreover, when the magnetic field is introduced, the vacuum energy acquires a Landau level (LL) structure and additional care is required in the treatment of the divergences. Many of the existing calculations of the properties of magnetized superconducting quark matter within this type of models remove these divergences by introducing some type of regulator function for each LL separately~\cite{Fayazbakhsh:2010gc,Fayazbakhsh:2010bh,Mandal:2012fq,Mandal:2016dzg,Mandal:2017ihr}. This procedure, however, might in general introduce unphysical oscillations. A discussion on this can be found in Refs.~\cite{Campanelli:2009sc,Frasca:2011zn,Gatto:2012sp}, where it is also observed that the use of smooth regulator functions improve the situation. In fact, this allows to identify possible physical oscillations appearing in some cases~\cite{Fukushima:2007fc,Noronha:2007wg}. However, an even clearer interpretation of the results can be obtained if the unphysical oscillations are completely removed with another scheme, especially at finite chemical potential and in the presence of color superconductivity. A regularization scheme of this type has been reported in Ref.~\cite{Menezes:2008qt} for the model in the absence of color superconductivity. The procedure follows the steps of the dimensional regularization prescription. This allows us to isolate the divergence into a term that has the form of the zero magnetic field vacuum energy and that can be regularized in the standard fashion. It should be stressed that similar expressions for the magnetic field dependent terms can be obtained using a method based on the proper-time formulation~\cite{Ebert:1999ht}. In a previous paper~\cite{Allen:2015paa}, some of the present authors used this scheme [so-called ``Magnetic Field Independent Regularization'' (MFIR) scheme] to study the influence of intense magnetic fields on the behavior of (isospin) symmetric color superconducting cold quark matter. The aim of the present work is to extend the corresponding results to the situation in which the conditions relevant for compact star applications are taken into account. Namely, our objective is to apply a NJL-type model within the MFIR scheme to study how the magnetic field affects the properties of cold dense two flavor quark matter under the constraints of color and electric charge neutrality as well as $\beta$-equilibrium. It is known that these constraints, which will be also referred to as compact star conditions, substantially modify the phase structure by suppressing some phases and favoring others~\cite{Buballa:2003qv,Alford:2007xm,Anglani:2013gfu,Shovkovy:2003uu, Huang:2003xd,Shovkovy:2004me}. As in the case of symmetric matter, the removal of unphysical oscillations induced by the regularization will allow us to construct and discuss the corresponding phase diagrams. 

This article is organized as follows. In Sec.~\ref{SecII} we present the NJL model with magnetic field and diquark interactions in the MFIR scheme for dense two flavor quark matter. We also impose compact star conditions in the limit of vanishing temperature. The model parameters used in our numerical calculations are also given. In Sec.~\ref{SecIII} we present our numerical results, discussing in detail the behavior of the different relevant quantities as functions of the magnetic field or the chemical potential. The corresponding phase diagrams are also presented and discussed. In Sec.~\ref{SecIV} we present our conclusions. Finally, in the Appendix we review some relevant features of the dependence of the zero magnetic field results on the strength of the quark-quark pairing interaction.

\section{The model and its regularization} \label{SecII}

\subsection{The thermodynamical potential in the mean field approximation}

In order to study the properties of cold dense two flavor quark matter under compact star conditions and in the presence of an external strong magnetic field, we consider the following Lagrangian density
\begin{equation}
\mathcal{L}=\mathcal{L}_{q}+\mathcal{L}_{lep}.
\label{lagr}
\end{equation}
The quark sector is described by a NJL-type $SU(2)_{f}$ Lagrangian density $\mathcal{L}_{q}$ which includes scalar-pseudoscalar and color pairing interactions and $\mathcal{L}_{lep}$ corresponds to the leptonic contribution. In the presence of an external magnetic field and finite chemical potentials, $\mathcal{L}_{q}$ reads
\begin{eqnarray}
\mathcal{L}_{q} & = & \bar{\psi}\left[i\ \tilde{\rlap/\!D}-m_{c}+\hat{\mu} \gamma^{0}\right]\psi\nonumber +G\left[\left(\bar{\psi}\psi\right)^{2}+\left(\bar{\psi}i\gamma_{5}\vec{\tau}\psi\right)^{2}\right] \\
 &  & +H\left[(i\bar{\psi}^{C} \epsilon_{f}\epsilon_{c}^{3}\gamma_{5}\psi)(i\bar{\psi} \epsilon_{f}\epsilon_{c}^{3}\gamma_{5}\psi^{C})\right].\label{Lq}
\end{eqnarray}
Here, $G$ and $H$ are coupling constants, $\psi=\left(u,d\right)^{T}$ represents a quark field with two flavors, $\psi^{C}=C\bar{\psi}^{T}$ and $\bar{\psi}^{C}=\psi^{T}C$, with $C=i\gamma^{2}\gamma^{0}$, are charge-conjugate spinors and $\vec{\tau}=(\tau_{1},\tau_{2},\tau_{3})$ are Pauli matrices. Moreover, $(\epsilon_{c}^{3})^{ab}=(\epsilon_{c})^{3ab}$ and $(\epsilon_{f})^{ij}$ are antisymmetric matrices in color and flavor space respectively. Furthermore, $m_{c}$ is the (current) quark mass that we take to be the same for both up and down flavors, and the diagonal chemical potential matrix $\hat{\mu}$ is $ \hat{\mu}=(\mu_{ur},\mu_{ug},\mu_{ub},\mu_{dr},\mu_{dg},\text{\ensuremath{\mu}}_{db})$, where the six quantities $\mu_{fc}$ are in principle independent parameters, but become related among themselves under compact star conditions.

In Eq.~(\ref{Lq}) we have introduced the covariant derivative $\tilde{D}_{\mu}=\partial_{\mu}-i\tilde{e}\tilde{Q}\tilde{{\cal A}}_{\mu}$. Note that here we are dealing with ``rotated'' fields. As is well known, in the presence of a non-vanishing superconducting gap $\Delta$, the photon acquires a finite mass. However, as shown in Ref.~\cite{Alford:1999pb}, there is a linear combination of the photon field $A_\mu$ and the eighth component of the gluon field $G_\mu^8$ that leads to a massless rotated $U(1)$ field $\tilde{A}_\mu$. The associated rotated charge matrix $\tilde{Q}$ is given by
\begin{equation}
\tilde{Q}=Q-\frac{T^8}{\sqrt{3}}=Q_{f}\otimes1_{c}-1_{f}\otimes\frac{T^8_c}{\sqrt{3}},
\end{equation}
where $Q_{f}=\mbox{diag}(2/3,-1/3)$ and $T^8_c=\mbox{diag}(1,1,-2)/2\sqrt{3}$. Then, in a six dimensional flavor-color representation $(ur,ug,ub,dr,dg,db)$, the rotated charges for the different quarks in units of $\tilde{e}$ are: $\tilde{q}_{ub}=1$, $\tilde{q}_{ur}=\tilde{q}_{ug}=1/2$, $\tilde{q}_{dr}=\tilde{q}_{dg}=-1/2$ and $\tilde{q}_{db}=0$. The rotated unit charge $\tilde{e}$ is given by $\tilde{e}=e\cos\theta$, where $\theta$ is the mixing angle which is estimated to be $\simeq1/20$~\cite{Gorbar:2000ms}. In the present work we consider a static and constant magnetic field in the 3-direction, $\tilde{{\cal A}}_{\mu}=\delta_{\mu2}x_{1}B$, which in fact is a mixture of the electromagnetic and color fields. Although here we are basically interested in the massless component of the rotated field, it is interesting to keep in mind that there also exists a massive X-component which can be either Meissner expelled or nucleated into vortices~\cite{Alford:2010qf}.

The leptonic contribution $\mathcal{L}_{lep}$ in Eq.~(\ref{lagr}) is given by the Dirac Lagrangian with chemical potential
\begin{equation}
\mathcal{L}_{lep}=\sum_l \  \bar{\psi}_{l}\left[i\gamma^{\mu}\left(\partial_{\mu}-ieA_{\mu}\right)-m_{l}+\text{\ensuremath{\mu}}_{l}\gamma^{0}\right]\psi_{l},
\end{equation}
where $l=e,\:\mu$ and we use $m_e=0.511\:\mathrm{MeV}$ and $m_\mu=105.66\:\mathrm{MeV}$. In order to describe the system as a function of $\tilde{e}B$, we will take $e\simeq \tilde{e}$ and $A_\mu \simeq \tilde{A}_\mu$ as a good approximation based on the small value of $\theta$.

The quark chemical potential matrix can furthermore be expressed in color-flavor space as $\hat{\mu}=\mu+Q\mu_{Q}+T^{8}\mu_{8}$ by introducing $\mu$, the common chemical potential for non-zero baryonic density, and the chemical potentials $\mu_{8}$ and $\mu_{Q}$, which are added to ensure color and electric charge neutrality conditions respectively. Here, we have used the fact that since the red and green quarks paired by the interaction are degenerate, their densities will be equal and we can impose $\mu_{3}=0$. Furthermore, assuming that no neutrinos are trapped in the system, $\beta$-equilibrium conditions lead to $\mu_{\mu}\!=\!\mu_{e}$ and $\mu_{dc}\!=\!\mu_{uc}+\mu_{e}$, where the latter implies $\mu_{Q}\!=\!-\mu_{e}$. While $\mu_{8}$ induces a difference between the chemical potentials of color paired (red and green) and unpaired (blue) quarks of the same flavor, $\mu_{e}$ differentiates the chemical potentials of flavored quarks ($u$ and $d$). Therefore, when 
$\beta$-equilibrium is taken into account, each chemical potential can be expressed in the following way:
\begin{equation}
\begin{aligned} & \mu_{ur}=\mu_{ug}=\mu-\frac{2}{3}\mu_{e}+\frac{1}{3}\mu_{8};\qquad\mu_{ub}=\mu-\frac{2}{3}\mu_{e}-\frac{2}{3}\mu_{8},\\
 & \mu_{dr}=\mu_{dg}=\mu+\frac{1}{3}\mu_{e}+\frac{1}{3}\mu_{8};\qquad\mu_{db}=\mu+\frac{1}{3}\mu_{e}-\frac{2}{3}\mu_{8}.
\end{aligned}
\label{relmu1}
\end{equation}
Again, the equality between chemical potentials of red and green quarks comes from the fact that the interaction pairs them in a degenerated way. For calculational simplicity, it is also convenient to define
\begin{equation}
\begin{aligned}\bar{\mu} & \equiv\frac{1}{2}\left(\mu_{dg}+\mu_{ur}\right)=\frac{1}{2}\left(\mu_{dr}+\mu_{ug}\right)=\mu-\frac{1}{6}\mu_{e}+\frac{1}{3}\mu_{8},\\
\delta\mu & \equiv\frac{1}{2}\left(\mu_{dg}-\mu_{ur}\right)=\frac{1}{2}\left(\mu_{dr}-\mu_{ug}\right)=\frac{1}{2}\mu_{e},
\end{aligned}
\label{relmu2}
\end{equation}
where the last equality in each line follows from $\beta$-equilibrium. 

In what follows we work in the mean field approximation (MFA), assuming that the only non-vanishing expectation values are $<\bar{\psi}\psi>=-(M-m_{c})/2G$ and $<i\bar{\psi}^{C} \epsilon_{f}\epsilon_{c}^{3}\gamma_{5}\psi>=-\Delta/2H$, which can be chosen to be real. Here, $M$ and $\Delta$ are the so-called dressed quark mass and superconducting gap, respectively. Following the standard procedure described in e.g. Refs.~\cite{Andersen:2014xxa,Mandal:2012fq}, the resulting MFA thermodynamic potential at vanishing temperature reads
\begin{equation}
\Omega_{\mbox{\scriptsize{MFA} }}=\frac{(M-m_{c})^{2}}{4G}+\frac{\Delta^{2}}{4H}-\sum_{\left|\tilde{q}\right|=0,\frac{1}{2},1}P_{\left|\tilde{q}\right|}-P_{lep},\label{omegaMFA}
\end{equation}
where $P_{lep}$ is the leptonic contribution to be explicitly given below (see Eq.~(\ref{P_lep})) and
\begin{eqnarray}
P_{\left|\tilde{q}\right|=0} & = & \int\frac{d^{3}p}{(2\pi)^{3}}\big(E_{db}^{+}+\left|E_{db}^{-}\right|\big),\label{omega0}\\
P_{\left|\tilde{q}\right|=1} & = & \frac{\tilde{e}B}{8\pi^{2}}\sum_{k=0}^{\infty}\alpha_{k}\int_{-\infty}^{\infty}dp_{z}\big(E_{ub}^{+}+\left|E_{ub}^{-}\right|\big),\label{omega1}\\
P_{\left|\tilde{q}\right|=1/2} & = & \frac{\tilde{e}B}{8\pi^{2}}\sum_{k=0}^{\infty}\alpha_{k}\int_{-\infty}^{\infty}dp_{z}\big(E_{\Delta^{+}}^{+}+|E_{\Delta^{-}}^{+}|+E_{\Delta^{+}}^{-}+|E_{\Delta^{-}}^{-}|\big).\label{omega12}
\end{eqnarray}
Here, we have introduced $\alpha_{k}=2-\delta_{k0}$ and
\begin{eqnarray}
E_{db\hphantom{_{+}}}^{\pm} & = & \sqrt{p^{2}+M^{2}}\pm\mu_{db}=E_{p}\pm\mu_{db},\\
E_{u_{b\hphantom{_{+}}}}^{\pm} & = & \sqrt{p_{z}^{2}+2k\tilde{e}B+M^{2}}\pm\mu_{ub}=E_{p_{z},k}\pm\mu_{ub}, \label{E1}\\
E_{\Delta_{\hphantom{_{+}}}^{\pm}}^{\pm} & = & \sqrt{\left(\sqrt{p_{z}^{2}+k\tilde{e}B+M^{2}}\pm\bar{\mu}\right)^{2}+\Delta^{2}}\pm\delta\mu=E_{\Delta,k}^{\pm}\pm\delta\mu.\label{E12}
\end{eqnarray}

Clearly, Eqs.~(\ref{omega0}-\ref{omega12}) are divergent and, thus, require to be regularized. In the next subsection we discuss the regularization scheme used to achieve this.

\subsection{MFIR regularization in the presence of pairing interactions and compact star conditions}

We start by considering the contributions corresponding to $P_{\left|\tilde{q}\right|=0,1}$. Since they do not involve the superconducting gap their treatment turns out to be simpler. In particular, $P_{\left|\tilde{q}\right|=0}$ does not depend explicitly on the magnetic field. Thus, it can be treated in the usual way~\cite{reports}. Namely, the expression in Eq.~(\ref{omega0}) can be separated into two terms
\begin{equation}
P_{\left|\tilde{q}\right|=0}=2\int\frac{d^{3}p}{(2\pi)^{3}}E_{p}+2\int\frac{d^{3}p}{(2\pi)^{3}}(\mu_{db}-E_{p})\Theta(\mu_{db}-E_{p}).\label{omegazero}
\end{equation}
The first term in this expression represents a vacuum contribution which turns out to be divergent. Typically, it can be regularized by introducing a cutoff function $h_{\Lambda}(p)$, which goes to zero for high momenta in order that the integral remains finite. The simplest choice is to take $h_{\Lambda}(p)=\Theta(\Lambda-p)$. In this case, we get
\begin{equation}
P_{VAC}^{reg}(M)=\frac{1}{\pi^{2}}\int_{0}^{\Lambda}dp\ p^{2}\ \sqrt{p^{2}+M^{2}}.\label{P_vac}
\end{equation}
On the other hand, the presence in the second term of Eq.~(\ref{omegazero}) of a chemical potential dependent Heaviside function makes it finite. Thus, no regularization is needed in this case. Explicitly, such term takes the form
\begin{equation}
P_{MED}(\mu_{db},M)=\frac{\Theta(\mu_{db}-M)}{\pi^{2}}\left[\frac{\mu_{db}(\mu_{db}^{2}-M^{2})^{3/2}}{3}-\frac{(\mu_{db}^{2}-M^{2})^{2}}{8}h\left(\frac{M}{\sqrt{\mu_{db}^{2}-M^{2}}}\right)\right], \label{P_med}
\end{equation}
where $h(z)=(2+z^{2})\sqrt{1+z^{2}}+z^{4}\ln[z/(1+\sqrt{1+z^{2}})]$. Consequently, the regularized form of $P_{\left|\tilde{q}\right|=0}$ reads
\begin{equation}
P_{\left|\tilde{q}\right|=0}^{reg}=P_{VAC}^{reg}(M)+P_{MED}(\mu_{db},M).
\end{equation}

The term $P_{\left|\tilde{q}\right|=1}$ is explicitly dependent on magnetic field. As in the previous case, it is convenient to separate it into two terms
\begin{equation}
P_{\left|\tilde{q}\right|=1}=\frac{\tilde{e}B}{4\pi^{2}}\sum_{k=0}^{\infty}\alpha_{k}\int_{-\infty}^{\infty}dp_{z}E_{p_{z},k}+\frac{\tilde{e}B}{4\pi^{2}}\sum_{k=0}^{\infty}\alpha_{k}\int_{-\infty}^{\infty}dp_{z}\big(\mu_{ub}-E_{p_{z},k}\big)\Theta\big(\mu_{ub}-E_{p_{z},k}\big).\label{P1}
\end{equation}
In this case the first contribution consists of an infinite sum of Landau level (LL) integrals, each of which is in turn divergent. As is discussed in the literature there are several possible regularization procedures (see e.g.~\cite{Duarte:2015ppa} and references therein). Here we perform the so-called ``Magnetic Field Independent Regularization'' (MFIR) described in~\cite{Menezes:2008qt}, in which the divergence is removed by subtracting a vacuum term with the form of the $\tilde eB=0$ case. This method has the advantage of removing spurious oscillations in the order parameters which show up in other procedures where each LL is regularized individually. Following the steps discussed in that reference, and taking into account that the only difference with respect to the no-pairing case is the rotated charge, we get the following replacement
\begin{equation}
\frac{\tilde{e}B}{4\pi^{2}}\sum_{k=0}^{\infty}\alpha_{k}\int_{-\infty}^{\infty}dp_{z}\ E_{p_{z},k}\rightarrow P_{VAC}^{reg}(M)+P_{MAG}(M,\tilde{e}B),\label{omegamag1}
\end{equation}
where $P_{VAC}^{reg}(M)$ was defined in Eq.~(\ref{P_vac}) and the (finite) vacuum magnetic term is
\begin{equation}
P_{MAG}(M,\tilde{e}B)=\frac{(\tilde{e}B)^{2}}{2\pi^{2}}\left[\zeta'(-1,x)+\frac{x-x^{2}}{2}\ \ln x+\frac{x^{2}}{4}\right]. \label{P_mag}
\end{equation}
Here, $\zeta(s,x)$ is the Hurwitz zeta function and $x=M^{2}/(2\tilde{e}B)$. Therefore, the divergence has been isolated into a vacuum term with the form of Eq.~(\ref{P_vac}), which is regularized with a sharp cut-off, while the explicit magnetic field dependence is contained in $P_{MAG}$. The medium term, which is finite, can be integrated explicitly once again, resulting in
\begin{equation}
P_{MED}(\mu_{ub},M,\tilde{e}B)=\Theta(\mu_{ub}-M)\frac{\tilde{e}B}{4\pi^{2}}\sum_{k=0}^{k_{max}}\alpha_{k}\left[\mu_{ub}\sqrt{\mu_{ub}^{2}-s_{k}^{2}}-s_{k}^{2}\ln\left(\frac{\mu_{ub}+\sqrt{\mu_{ub}^{2}-s_{k}^{2}}}{s_{k}}\right)\right],\label{P_med_eB}
\end{equation}
where $k_{max}=\mathrm{Int}\left[ (\mu_{ub}^{2}-M^{2})/(2\tilde{e}B) \right] $ and $s_{k}=\sqrt{M^{2}+2k\tilde{e}B}$. Gathering these expressions we get
\begin{equation}
P^{reg}_{\left|\tilde{q}\right|=1}=P^{reg}_{VAC}(M)+P_{MAG}(M,\tilde{e}B)+P_{MED}(\mu_{ub},M,\tilde{e}B).\label{omegaone}
\end{equation}

Finally, we treat the case of the quark species with ${\left|\tilde{q}\right|=1/2}$ which corresponds to the paired quarks and is therefore more involved. First, we separate the contributions including the parameter $\delta\mu$, much in the same way in which the medium term is separated from the vacuum term in Eq.~(\ref{P1}). The resulting expression is
\begin{eqnarray}
P_{\left|\tilde{q}\right|=1/2}
&=&P_{\Delta}(M,\Delta, \bar \mu,\tilde{e}B) + P_{\delta\mu}(M,\Delta, \bar \mu, \delta \mu,\tilde{e}B),
\label{omegahalf}
\end{eqnarray}
where
\begin{eqnarray}
P_{\Delta}(M,\Delta, \bar \mu,\tilde{e}B) &= & 
\frac{\tilde{e}B}{4\pi^{2}}\sum_{k=0}^{\infty}\alpha_{k}\int_{-\infty}^{\infty}dp_{z}\sum_{s=\pm} E_{\Delta,k}^{s} \ ,
\\
P_{\delta\mu}(M,\Delta, \bar \mu, \delta \mu,\tilde{e}B) &=&
\frac{\tilde{e}B}{4\pi^{2}}\sum_{k=0}^{\infty}\alpha_{k}\int_{-\infty}^{\infty}dp_{z}\sum_{s=\pm}
(\delta\mu-E_{\Delta,k}^{s})\Theta(\delta\mu-E_{\Delta,k}^{s}).
\end{eqnarray}
Next, noting that the first term in Eq.~(\ref{omegahalf}) is divergent, we regularize it following the steps in the Appendix of Ref.~\cite{Allen:2015paa}. We obtain
\begin{eqnarray}
P^{reg}_{\Delta}(M,\Delta, \bar \mu,\tilde{e}B)& = & \frac{2}{\pi^{2}}\int_{0}^{\Lambda}dp\ p^{2}\ \left(E_{\Delta}^{+}+E_{\Delta}^{-}\right)+\frac{(\tilde{e}B)^{2}}{2\pi^{2}}\left[\xi'(-1,y)+\frac{y-y^{2}}{2}\ln y+\frac{y^{2}}{4}\right]\nonumber \\
 &  & +\frac{(\tilde{e}B)^{2}}{2\pi^{2}}\int_{0}^{\infty}dp\ \left[\sum_{k=0}^{\infty}\alpha_{k}\ f(p^{2}+k)-2\int_{0}^{\infty}dt\ f(p^{2}+t)\right],\label{P_Delta}
\end{eqnarray}
where $E_{\Delta}^{\pm}=\sqrt{(\sqrt{p^{2}+M^{2}}\pm\bar{\mu})^{2}+\Delta^{2}}$,
$y=(M^{2}+\Delta^{2})/(\tilde{e}B)$ and
\begin{equation}
f(z)=\sum_{s=\pm1}\left[\sqrt{(\sqrt{z+2x}+s\:\bar{\mu}/\sqrt{\tilde{e}B})^{2}+y-2x}-\sqrt{z+y}\right].
\end{equation}
As in previous cases, the first term in Eq.~(\ref{P_Delta}) has been regularized with a sharp cutoff. This term contains contributions from vacuum and matter which do not explicitly depend on the magnetic field, and cannot be disentangled into two separate terms unless $\Delta=0$. The second term is the vacuum magnetic contribution which was also found in the $|\tilde{q}|=1$ case, now generalized to the case $\Delta\neq0$. Finally, the third term is an additional explicitly magnetic field dependent matter contribution which is finite, as proved in Ref.~\cite{Allen:2015paa}.

Turning finally to the second term in Eq.~(\ref{omegahalf}), we note that some conditions have to be satisfied for this term to be non-zero. This is more clearly seen when $P_{\delta\mu}$ is re-written in the following form
\begin{equation}
P_{\delta\mu}(M,\Delta, \bar \mu, \delta \mu,\tilde{e}B)=\Theta(\delta\mu-\Delta)\frac{\tilde{e}B}{2\pi^{2}}\sum_{s=\pm}s\:\Theta(k_{s})\sum_{k=0}^{\mathrm{Int}[ k_{s}]}\alpha_{k}\int_{0}^{p_{s}}dp_{z}(\delta\mu-E_{\Delta,k}^{-}), \label{P_deltamu}
\end{equation}
where $k_{\pm}=\left[(\bar{\mu}\pm\sqrt{\delta\mu^{2}-\Delta^{2}})^{2}-M^{2}\right]/\tilde{e}B$ and $p_{\pm}=\sqrt{(k_{\pm}-k)\tilde{e}B}$. The first Heaviside function serves to distinguish between two possible situations within a $\Delta \neq 0$ phase: a ``gapless phase'' ($g2SC$) when $\delta\mu>\Delta$, and an ordinary two color superconducting phase ($2SC$) when $\Delta>\delta\mu$ and this term vanishes. The source of the differences between these phases comes from the changes in the quasi-particle spectrum in Eq.~(\ref{E12}). As explained in Refs.~\cite{Huang:2003xd,Shovkovy:2004me}, when $\delta\mu\neq0$ the gap equation has two branches of solutions and the modes are no longer completely degenerate, but they split into pairs of two with gaps $\Delta_{\pm}=\Delta\pm\delta\mu$. While in the $2SC$ phase the four modes are gapped, in the $g2SC$ phase the lower dispersion relation for the quasi-particle crosses the zero-energy axis and two of the four modes become gapless.

Having discussed the regularization of the quarks contribution to Eq.~(\ref{lagr}), only the leptonic term $P_{lep}$ remains to be considered. However, since leptons also have unit charge, the total leptonic pressure is quickly recovered from  $P_{\left|\tilde{q}\right|=1}$ upon performing the replacements $M \rightarrow m_l$ and $\mu_{ub}\rightarrow \mu_e$. Since $m_{l}$ and $B$ are fixed, the vacuum and magnetic terms can be ignored when one looks for the minimum of the potential and we have
\begin{equation}
P_{lep} = \sum_{l=e,\mu} P_{MED}(\mu_{e},m_l,\tilde{e}B).\label{P_lep}
\end{equation}

Therefore, as a result of the regularization procedure we finally get
\begin{eqnarray}
\Omega_{\mbox{\scriptsize{MFA}}}^{reg}&=&\frac{(M-m_{c})^{2}}{4G}+\frac{\Delta^{2}}{4H}-2P_{VAC}^{reg}(M)-P_{MED}(\mu_{db},M) \nonumber \\
&& -P_{MAG}(M,\tilde{e}B)-P_{MED}(\mu_{ub},M,\tilde{e}B)-\sum_{l=e,\mu} P_{MED}(\mu_e,m_l,\tilde{e}B) \nonumber \\
&& -P_{\Delta}^{reg}(M,\Delta, \bar \mu,\tilde{e}B) - P_{\delta\mu}(M,\Delta, \bar \mu, \delta \mu,\tilde{e}B),\label{omegaMFAreg}
\end{eqnarray}
where the corresponding terms have been defined in Eqs.~(\ref{P_vac},\ref{P_med},\ref{P_mag},\ref{P_med_eB},\ref{P_Delta},\ref{P_deltamu}). Given this form for $\Omega_{\mbox{\scriptsize{MFA}}}^{reg}$, the minimum for fixed values of $\mu$ and $\tilde{e}B$ is found by solving the gap equations
\begin{equation}
\frac{\partial\Omega_{\mbox{\scriptsize{MFA}}}^{reg}}{\partial\xi}=0,\qquad\xi=M,\Delta,\label{gapeq}
\end{equation}
subject to the neutrality and $\beta$-equilibrium conditions. Neutrality for electric and color charge can be
expressed in the following form
\begin{equation}
\frac{2}{3}n_{u}-\frac{1}{3}n_{d}=n_{e}+n_{\mu} \label{neutrality1a} 
\end{equation}
\begin{equation}
n_{r}=n_{g}=n_{b}. \label{neutrality1b}
\end{equation}
In these equations, we have introduced
\begin{equation}
n_f = \sum_{c} n_{fc},\qquad n_c = \sum_{f} n_{fc},
\end{equation}
where the density for each quark (lepton) species is obtained by deriving the thermodynamical potential with respect to the corresponding quark (lepton) chemical potential $\mu_{fc}$ ($\mu_l$). For this purpose, the 
chemical potentials of the eight particle species must be treated as independent variables. Then, the system consisting of Eqs.~(\ref{gapeq},\ref{neutrality1a},\ref{neutrality1b}) must be solved for $M$, $\Delta$, $\mu_e$ and $\mu_8$, where the relations (\ref{relmu1}) are to be applied to express all chemical potentials in terms of $\mu$, $\mu_e$ and $\mu_8$ (it must be taken into account that since we have already taken $\mu_3=0$, the equation $n_r=n_g$ is automatically satisfied, hence (\ref{neutrality1b}) is actually only one equation). For each value of $\mu$ and $\tilde{e}B$, several solutions of these equations will generally exist, corresponding to different possible phases. In particular there is a larger amount of solutions with respect to the zero magnetic field case due to the fact that there can be different solutions with different values of $k_{max}$. The most stable solution is that associated with the absolute minimum of the thermodynamic potential.

Finally, it is interesting to note that, on imposing $\beta$-equilibrium conditions and re-expressing the chemical potentials in terms of $\mu$, $\mu_e$ and $\mu_8$, the neutrality conditions (\ref{neutrality1a}) and (\ref{neutrality1b}) can be written in the alternative form
\begin{equation}
\frac{\partial\Omega^{reg}_{\mbox{\scriptsize{MFA}}}}{\partial\mu_{e}}=0,\qquad\frac{\partial\Omega^{reg}_{\mbox{\scriptsize{MFA}}}}{\partial\mu_{8}}=0.\label{neutrality2}
\end{equation}

For future reference, it is useful at this point to comment on some details concerning the densities of the different quark species.
For the unpaired species, that is, the $b$-quarks, they are given by
\begin{eqnarray}
n_{db} &=& \Theta(\mu_{db}-M) \frac{1}{3 \pi^2} ( \mu^2_{db} - M^2)^{3/2}, \label{n_db}\\
n_{ub} &=& \Theta(\mu_{ub}-M) \frac{1}{2 \pi^2} \sum_{k=0}^{k_{max}} \alpha_k |\tilde {e} B| \sqrt {\mu_{ub}^2 - M^2 - 2 k |\tilde {e} B|}.\label{n_ub}
\end{eqnarray}
The lepton densities can be obtained from $n_{ub}$ by replacing $\mu_{ub}\!\rightarrow\! \mu_e$  and $M\!\rightarrow\! m_l$.
The explicit expressions for the paired quarks densities are more complicated. They cannot be obtained from the thermodynamical potential as written in Eq.~(\ref{omegaMFAreg}) since it is already evaluated at $\mu_3\!=\!0$. We can however relax this condition through the replacement 
\begin{equation}
P^{reg}_\Delta (M,\Delta,\bar \mu, \tilde eB) \rightarrow \frac{1}{2}  P_\Delta^{reg} \left(M,\Delta, \frac{\mu_{ur} + \mu_{dg}}{2}, \tilde eB \right) + \frac{1}{2} P_\Delta^{reg} \left(M,\Delta,\frac{\mu_{ug}+\mu_{dr}}{2}, \tilde eB \right), 
\end{equation}
where a similar one must be made for $P_{\delta \mu}(M,\Delta,\bar \mu,\delta \mu,\tilde eB)$. Deriving the obtained expression with respect to $\mu_{fc}$ and evaluating at $\mu_3=0$ afterwards, we obtain: 
\begin{equation}
n_{fc}= \frac{1}{4}\left( \frac{\partial P^{reg}_{\Delta}}{\partial \bar \mu} + \frac{\partial P_{\delta \mu}}{\partial \bar \mu} \pm
\frac{\partial P_{\delta \mu}}{\partial (\delta\mu)} \right) \qquad \mbox{for c=r,g},
\end{equation}
where the plus (minus) sign corresponds to down (up) quarks and $n_{fr}=n_{fg}$. 

While the behavior of the $db$ quarks is quite simple, we see that for $ub$ quarks, which interact with the magnetic field, $n_{ub}$ contains a sum up to a $k_{max}$ value which is determined by $\mu_{ub}$, $M$ and $\tilde {e} B$. These quarks will occupy all LL's up to this $k_{max}$ number. The leptonic densities are analogous. On the other hand, the sum over LL's inside the $P_\Delta$ term is infinite, which means that all LL's are occupied for the paired quarks. Furthermore, when $P_{\delta \mu}$ is finite the $u$ and $d$ densities of the paired quarks will not be equal. It is interesting to note that $n_{ur}=n_{ug}$ and $n_{dr}=n_{dg}$, since they have identical chemical potentials, and
\begin{equation}
n_{dr}-n_{ur}=n_{dg}-n_{ug}=\frac{1}{2}\frac{\partial P_{\delta\mu}}{\partial (\delta\mu)}.
\label{densrels}
\end{equation}
While in the $2SC$ phase $P_{\delta\mu}=0$ and the densities of the quark species that participate in pairing dynamics are equal, they can get separated in the $g2SC$ phase.

\subsection{Model parametrization\label{subsec:II.B}}

In order to analyze the dependence of the results on the model parameters, we will consider two $SU(2)_{f}$ NJL model parametrizations. Set~1 leads to $M_{0}=340\:\mathrm{MeV}$ while Set~2 to $M_{0}=400\;\mathrm{MeV}$. Here, $M_{0}$ represents the vacuum effective quark mass in the absence of external magnetic fields. The corresponding model parameters are listed in Table~\ref{pnjl}.

\begin{table}[h]
\begin{centering}
\caption{\label{pnjl} Parameter sets for the SU(2)$_{f}$ NJL model. In both
cases, empirical values in vacuum for the pion observables are reproduced,
$m_{\pi}=138\:\mathrm{MeV}$ and $f_{\pi}=92.4\:\mathrm{MeV}$.}
\vspace{0.5cm}
\par\end{centering}
\centering{}%
\begin{tabular}{cccccc}
\hline
\hline\hspace*{0.2cm} Parameter set \hspace*{0.2cm}  & \hspace*{0.2cm} $M_{0}$ \hspace*{0.2cm} & \hspace*{0.2cm} $m_{c}$ \hspace*{0.2cm}  & \hspace*{0.2cm} $G\Lambda^{2}$ \hspace*{0.2cm}  & \hspace*{0.2cm} $\Lambda$ \hspace*{0.2cm}  & \hspace*{0.2cm} $-<u\bar{u}>^{1/3}$ \hspace*{0.2cm} \tabularnewline
 & MeV  & MeV  &  & MeV  & MeV \tabularnewline
\hline
Set~1  & 340  & 5.59  & 2.21  & 621  & 244 \tabularnewline
Set~2  & 400  & 5.83  & 2.44  & 588  & 241 \tabularnewline
\hline
\end{tabular}
\end{table}

\section{Numerical Results} \label{SecIII}

In this section we present our results for the properties of cold color superconducting quark matter subject to compact star conditions under the influence of an external magnetic field. We will carry out a detailed study of the order parameters as a function of the model parameters and, in particular, the coupling constant ratio $H/G$. The corresponding phase diagrams for both Set~1 and Set~2 will be presented as well. Concerning the strength of the diquark interaction, it should be mentioned that although the ratio $H/G\!=\!0.75$ is favored by various models of the quark effective interaction, from a more phenomenological point of view this value is subject to rather large uncertainties~\cite{Buballa:2003qv}. Thus, here we will consider the representative values $H/G\!=\!0.75$ and $1$, which in fact give rise to different possible phase structures of the system, as will be seen. The situation corresponding to other values of $H/G$ will be briefly addressed.

\subsection{Order parameters as a function of the chemical potential}
\label{tresuno}

\begin{figure}
\centering{}\includegraphics[width=1\textwidth]{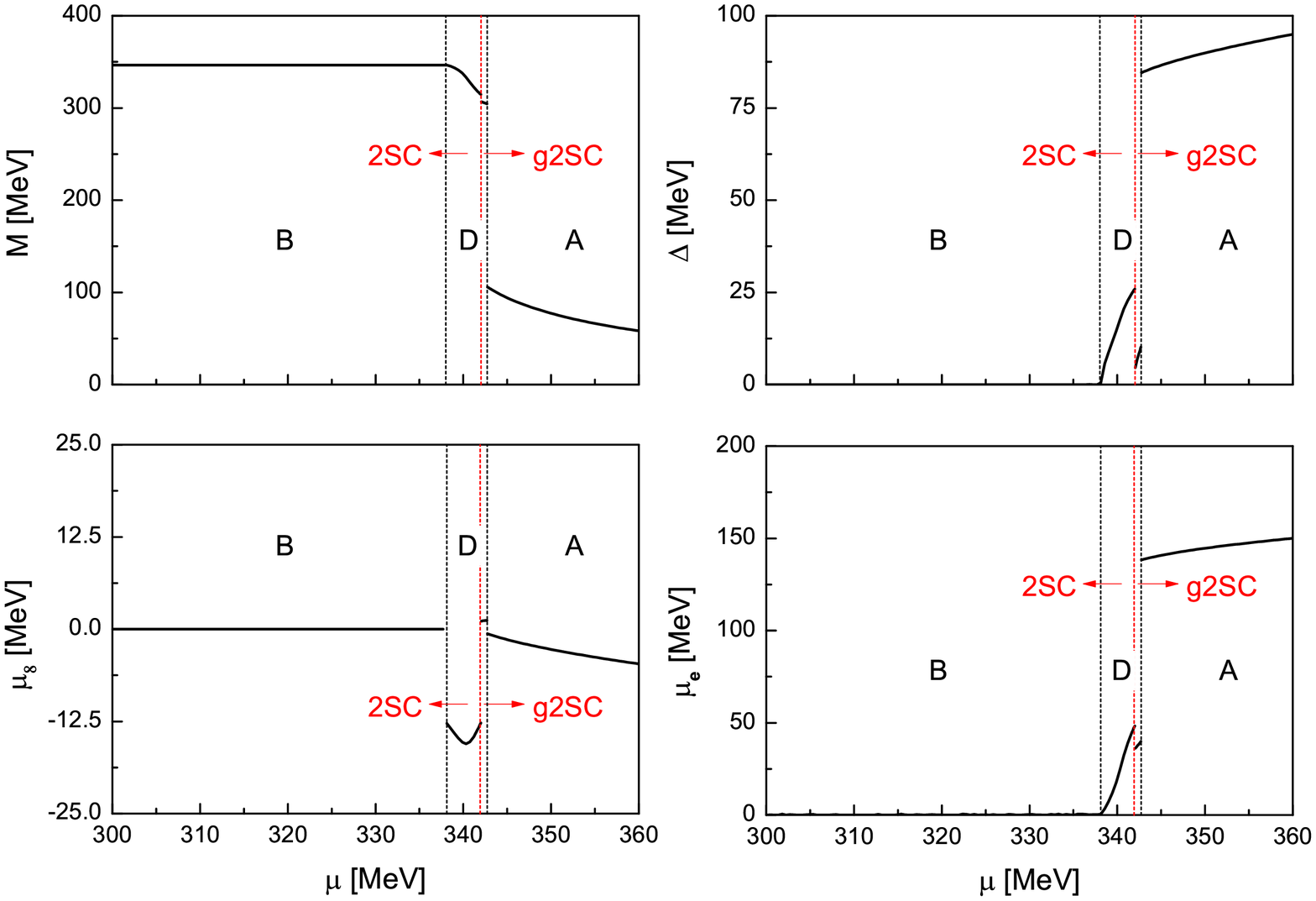}
\caption{(Color online) $M$, $\Delta$, $\mu_{8}$ and $\mu_{e}$ vs $\mu$ for $H/G=0.75$ and $\tilde{e}B=0.07\:\mathrm{GeV^2}$, Set~1. Black dotted vertical lines separate $B$-$D$-$A$ phases while the red ones separate $2SC$-$g2SC$ phases.}
\label{Fig1}
\end{figure}

In this subsection we analyze the behavior of the order parameters as a function of $\mu$ for given values of the magnetic field. We will refer mainly to the results for Set~1 since they exhibit a richer structure. Given the potential complexity of the phase structure, it is convenient to start with a presentation of the results for a particular value of $\tilde eB$, which we display in Fig.~\ref{Fig1}. Here, we describe the main features of the phases present in the model and introduce the corresponding notation, following the one used in previous studies~\cite{Allen:2015paa,Ebert:1999ht,Ebert:2003yk,Allen:2013lda,Allen:2015qxa,Grunfeld:2014qfa}. The massive, vacuum and non-superconducting phase which is present for lower chemical potentials in the displayed figures is denoted as the $B$-phase. On the other hand, the (almost) chirally restored and $\Delta \neq 0$ phase appearing for the larger chemical potentials is denoted as the $A$-phase. Depending on the model parameters and magnetic field, a phase which we refer to as a $D$-phase may also be present at intermediate chemical potentials, in which both the chiral and superconducting order parameters are finite and fairly large. For this reason, this phase is sometimes referred to as a ``mixed phase''~\cite{Huang:2002zd}, even though this is not the only meaning which has been given to it in the literature~\cite{Neumann:2002jm}. Furthermore, the superconducting phases $D$ and $A$ can in turn find themselves in two possible modes, depending on the relative values of $\delta \mu$ and $\Delta$. When $\delta \mu > \Delta$, the associated phase is said to be gapless (g2SC), and in the opposite case the phase is simply referred to as a $2SC$ type phase. 

\begin{figure}
\centering{}\includegraphics[height=0.9\textheight]{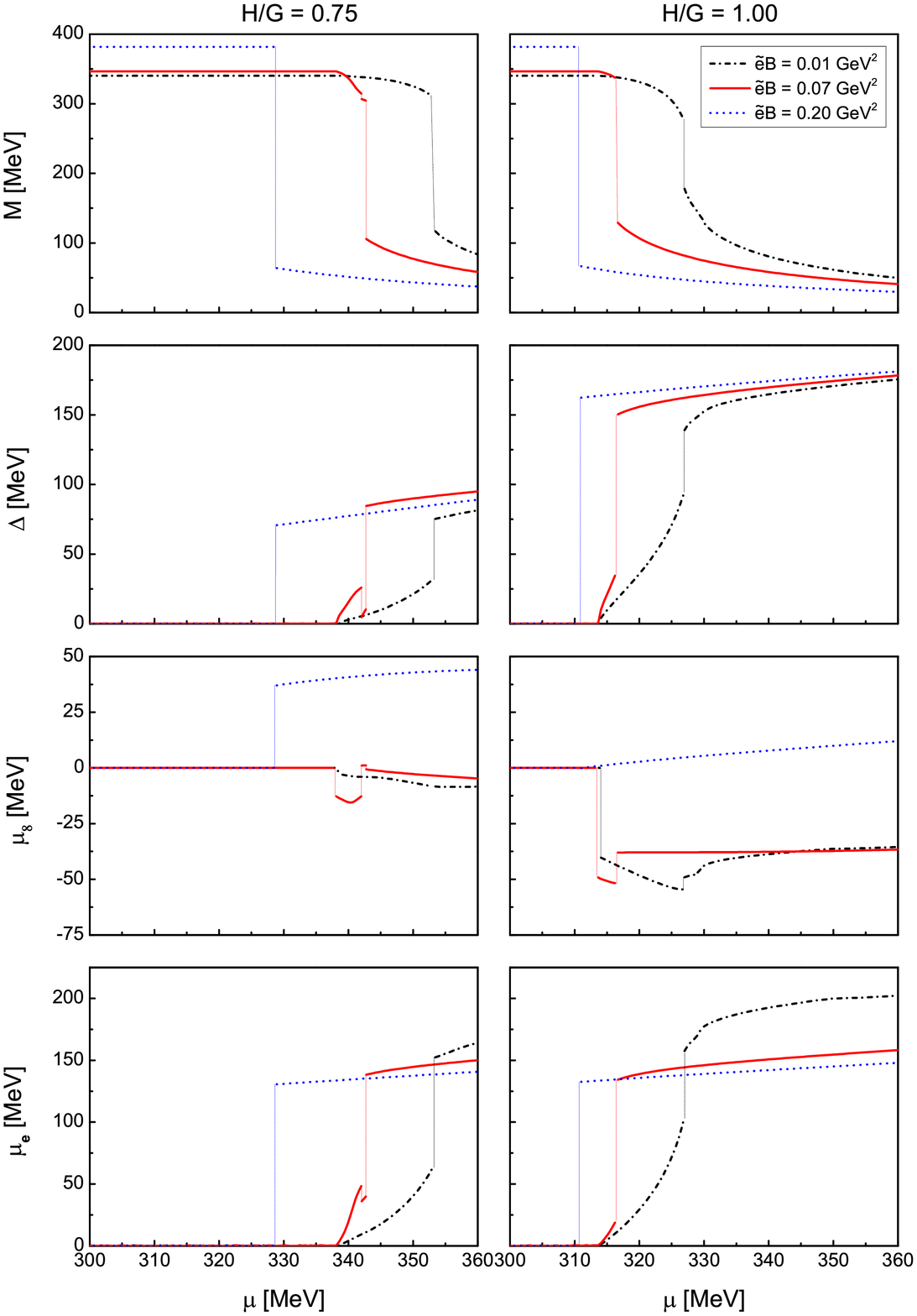}
\caption{(Color online) $M$, $\Delta$, $\mu_{8}$ and $\mu_{e}$ vs $\mu$ for the two values of $H/G$ considered and three representative values of $\tilde{e}B$, Set~1.}
\label{Fig2}
\end{figure}

In Fig.~\ref{Fig2} we plot the results for $M$, $\Delta$, $\mu_e$ and $\mu_8$ as a function of $\mu$ for three representative values of $\tilde{e}B$. The $B$ phase is always found for low enough chemical potentials.
Here, the superconducting gap is zero and $\mu$ is not high enough to populate the quark and lepton species. As a matter of fact, this phase bears no difference with respect to the vacuum phase found in the case without diquark interactions, except for the fact that the magnetic field that modifies the constituent mass is rotated. The order parameter $M$ is independent of $\mu$ in this phase because the corresponding $\mu$-dependent terms in the thermodynamic potential vanish. In the vanishing $\tilde eB$ limit, $M$ takes the value $M_0$ determined by the set of parameters and it increases with the magnetic field, as expected from the magnetic catalysis effect in vacuum~\cite{Preis:2012fh,Shovkovy:2012zn}. Furthermore, since the vacuum is electric and color charge neutral, both $\mu_e$ and $\mu_8$ can be taken to be zero. For later reference we note here that, strictly speaking, there is a finite range of values for $\mu_8$ and $\mu_e$ that also lead to the vacuum solution. We choose to set $\mu_e=0$ and $\mu_8=0$ for simplicity.

For a given magnetic field, the system always finds itself in the $A$ phase for large enough chemical potentials. It is characterized by a large value of $\Delta$ and the property that chiral symmetry is restored. However, due to the presence of a finite current quark mass, such restoration will only be approximate. Near the transition, $M$ will take values around $100\:\mathrm{MeV}$ and diminish toward a value slightly above $m_c$ for higher chemical potentials. Since $\mu$ is well above the dressed mass, all quark species will be populated, so $\mu_8$ and $\mu_e$ will take non-zero values in order to enforce the neutrality conditions. When, for example, $\mu_8$ is negative we have that $\mu_{fp} < \mu_{fb}$ for $p=r,g$. In order to maintain a zero net color charge, an excess of blue quark density is generated with respect to the case without neutrality conditions. In the cases displayed in Fig.~\ref{Fig2}, $\mu_8$ is negative for the two lower values of the magnetic field, and can vary down to $-40$ MeV. In general, it will lie in the range $\left|\mu_{8}\right|<70\:\mathrm{MeV}$  and, as can be seen for example for $\tilde eB\!=\!0.20\:\mathrm{GeV^2}$, it can also take positive values for higher magnetic fields. As for $\Delta$ and $\mu_{e}$, they increase with $\mu$ in the range considered and can both acquire high values, of the order of $150\:\mathrm{MeV }$. When the $B$ and $A$ phases connect directly, they do so through a first order transition, as observed in Fig.~\ref{Fig2} for $\tilde{e}B\!=\!0.20\:\mathrm{GeV}^2$.

It should be noted that in all cases once the system is in the $A$ phase the order parameters display some tiny features (sometimes even hardly visible in the figures) at certain values of the chemical potential. They correspond to the so-called ``van Alphen-de Haas''(vA-dH) effect~\cite{Ebert:1999ht} and will be discussed in detail in the following subsections.

Finally, the $D$ phase is present for the two lower values of $\tilde{e}B$ in Fig.~\ref{Fig2}. In this phase, $M$ is lower than in vacuum but typically much larger than in the $A$ phase, hence chiral symmetry can be said to be only partially restored. The parameters $\Delta$ and $\mu_{e}$ increase from zero, acquiring rather large values in a short range of $\mu$. In fact, since $M$ and $\Delta$ change continuously when going from $B$ to $D$, the corresponding transition is of the second order type. We should also note that $\mu_8$ appears to be discontinuous along this transition for both $H/G$ values. However, we bear in mind that in the $B$ phase its value is actually not well-defined, and since it is arbitrarily taken to be zero, such discontinuity has no physical meaning. In the $D$ phase, it can actually occur that there is quark population for a given species even when $\mu < M$, which means that $H/G$ is large enough to dynamically break the symmetry of the color gauge group even at chemical potential and dressed mass values that would correspond to vacuum in the no-pairing case. The diquark interaction hence induces a density of $r$ and $g$ quarks. In order to satisfy color charge neutrality, blue quarks must be present as well, so $\mu_8$ acquires large and negative values such that $\mu_{fb}>M$, for at least one of the two flavors. The $D$ phase is connected to the $A$ phase through a first order transition. When going from $D$ to $A$, $\mu_8$ is still negative but takes a smaller absolute value. Both $\Delta$ and $\mu_{e}$ jump to larger values, but the rate of increase with respect to $\mu$ is smaller. Also, comparing the left and right panels we see that increasing the diquark interaction will always induce a larger gap in both the $A$ and $D$ phases. It is important to note that superconductivity is suppressed with respect to the non-neutrality case, where the $D$ phase is only present for larger values of $H/G$ as was noted in~\cite{Allen:2015paa}. It can be interpreted that, due to this suppression, the $D$ phase exists because the $A$ phase with larger $\Delta$ is energetically disfavored with respect to the former. The origin of the suppression lies in the fact that the presence of $\mu_e$ separates the Fermi momenta of the up and down quarks with respect to each other. Since the quark pairing occurs between particles of equal and opposite momenta, this splitting reduces the diquark condensate.

For the particular case $H/G=0.75$ and $\tilde{e}B=0.07\;\mathrm{GeV^{2}}$, we see that immediately after
the second order transition from $B$ to $D$, $\Delta\gtrsim\delta\mu$, so the system finds itself in the $2SC$ mode. However, these two quantities are very similar, and when the chemical potential is increased $\delta \mu$ becomes larger than $\Delta$ for $\mu\simeq342\;\mathrm{MeV}$, leading to a $g2SC$ region. We see that when the transition occurs, $M$ and $\mu_e$ diminish slightly while $\Delta$ and $\mu_8$ drop sharply, where the latter also changes sign. This phase exists in a very short $\mu$ range, so another first order transition occurs almost immediately to the $A$ phase.

\begin{figure}
\centering{}\includegraphics[height=0.9\textheight]{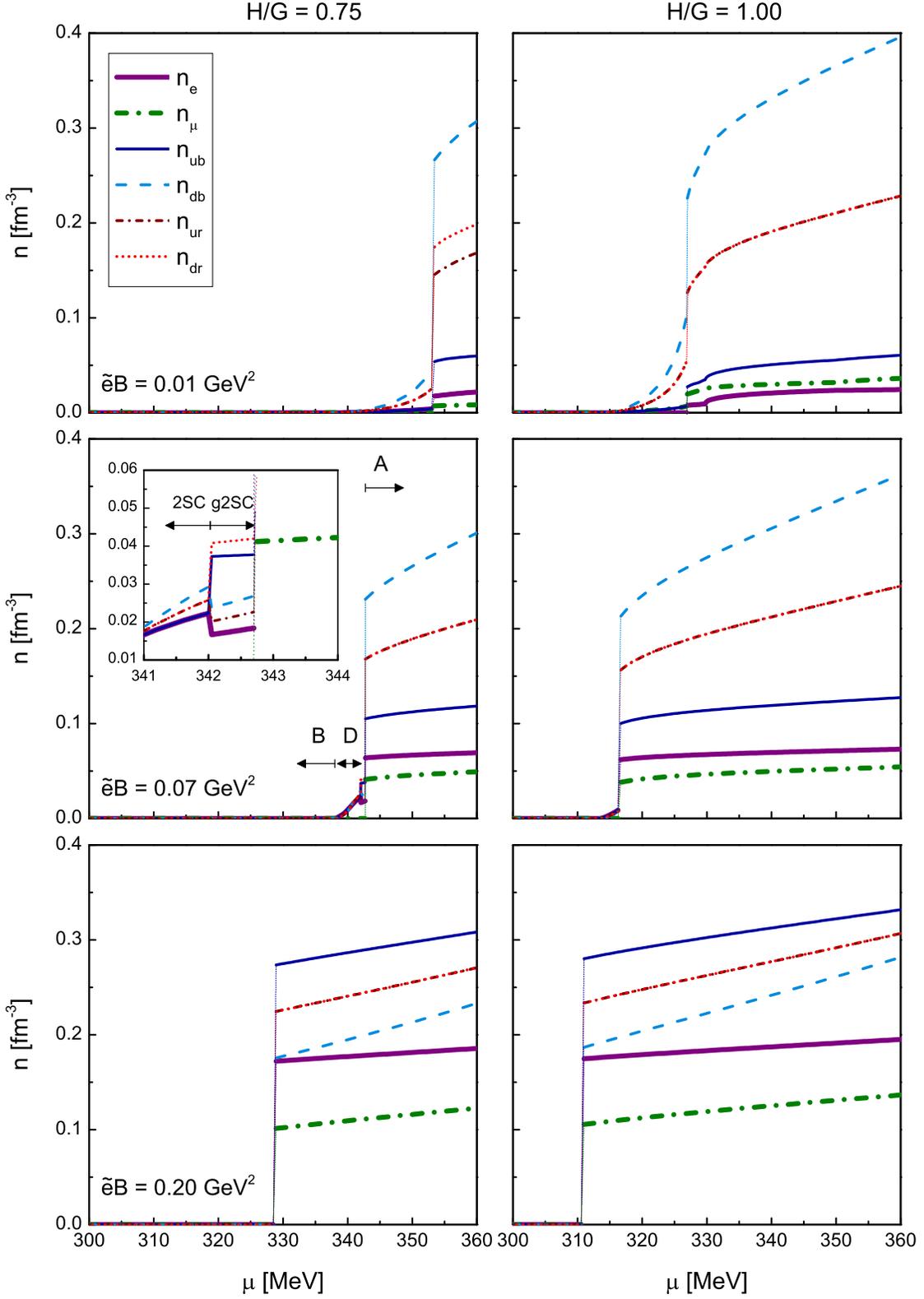}
\caption{(Color online) Quarks and leptons densities vs $\mu$ for the two values of $H/G$ considered and three representative values of $\tilde{e}B$, Set~1. The labels in the intermediate left panel indicate the different phases described in Fig~\ref{Fig1}.}
\label{Fig3}
\end{figure}

In Fig.~\ref{Fig3} we illustrate the corresponding densities which, of course, satisfy the neutrality conditions in Eqs.~(\ref{neutrality1a},\ref{neutrality1b}). We recall that due to Eq.~(\ref{relmu1}), one has $n_{fr}=n_{fg}$. Thus, of these four
densities, only those corresponding to the red quarks are given. As expected, all densities are zero in the $B$ phase, while in the $A$ phase the chemical potential is well above the quark dressed mass, therefore resulting in finite densities for all quarks. Furthermore, the lepton chemical potential is above 120 MeV, so both muon and electron population will be finite. As stated before, in the $D$ phase one finds quark population even for $\mu<M$. The densities of the $u$ and $d$ quarks forming diquark pairs are equal in the $2SC$ region because $P_{\delta \mu}=0$. On
the other hand, we observe in the inset of the intermediate left panel that $n_{ur}$ and $n_{dr}$ separate in the $g2SC$ region according to Eq.~(\ref{densrels}), since $P_{\delta \mu} \neq 0$. When the $D$ phase begins, $\Delta$ becomes positive populating the densities of the paired quarks, while the blue quarks also acquire a non-zero density to ensure neutrality. Because of electric charge neutrality, $\mu_e$ becomes positive. Electron population will therefore be finite, as a very small value of $\mu_e$ is enough to excite the corresponding lowest Landau level (LLL). As for the muons, since a higher $\mu_e$ is required to overcome $m_\mu$, the condition for their population does not get satisfied. Since $\mu_e\neq 0$ makes $\mu_{ub}$ lower than $\mu_{db}$, the density $n_{db}$ will be finite immediately after the transition, while $n_{ub}$ will be null until a value of $\mu$ slightly larger (barely visible in Fig.~\ref{Fig3}).

We end this subsection by briefly mentioning the situation concerning Set~2. For this set of parameters when one starts from the $B$ phase and steadily increases $\mu$ there is always a certain critical chemical potential at which the system undergoes a first order phase transition to an $A$ phase in the 2SC mode. Therefore, the behavior of the order parameters and densities is basically smooth except for the discontinuity at the critical $\mu$ and the quite small features related to the vA-dH effect.

\subsection{Phase diagrams in the $\boldsymbol{\tilde{e}B-\mu}$ plane}

After having described the behavior of the order parameters, we consider in this subsection the corresponding phase diagrams in the $\tilde{e}B-\mu$ plane. They are given in Fig.~\ref{Fig4} for the two coupling ratios and parameter sets considered. Comparing the top and bottom panels, we see that as the $H/G$ ratio is increased all the transition lines in the diagram are brought downwards: for a more intense pairing, a lower chemical potential is necessary to produce the superconducting phases. Regarding the left and right panels, we observe that Set~1 exhibits a more complex structure while in Set~2 there are less phases. The fact that the phase diagram is simplified when the parameters are modified such that $M_0$ is increased is in agreement with previous results~\cite{Allen:2015paa,Allen:2015qxa,Grunfeld:2014qfa}.
Furthermore, since Set~2 corresponds to a higher value of $M_0$ the transitions are displaced to higher chemical potentials. As mentioned above, in Set~2 only phases $B$ and $A$ exist, the last one being in the $2SC$ mode in its whole range. In Set~1, instead, phases $B$, $D$ and $A$ are present, where the latter two can exist both in the $g2SC$ or the $2SC$ modes. We will concentrate on Set~1 in what follows. As mentioned, the intermediate $D$ phase connects to the $B$ phase through a second order transition and to the upper $A$ phase through a first order transition, while $A$
and $B$ connect directly through a first order transition. The behavior as a function of $\tilde eB$ of the first order transition leading to the $A$-type phases is worth noting: for small fields, the transition has a small downward slope, which becomes sharper as the magnetic field is increased. The critical chemical potential reaches a minimum around $\tilde{e}B\simeq0.15-0.25\;\mathrm{GeV^{2}}$, after which it increases indefinitely with $\tilde eB$, therefore forming a well-shaped curve. This effect is related to the so-called ``inverse magnetic catalysis'' (IMC) effect~\cite{Preis:2012fh,Shovkovy:2012zn} in that, in a certain range of chemical potentials, an increase of the magnetic field at intermediate values favors the chirally restored phases. Therefore, we refer to this curve as the ``IMC well". It is interesting to note that when we increase $H/G$, the depth of the IMC well diminishes and its width increases. Also, compact star conditions tend to decrease the IMC effect: the depth of the well decreases, in agreement with what is observed in~\cite{Allen:2015qxa} for the case with vector interactions and without superconductivity.

The existence of the $D$ phase is a consequence of the diquark pairing alone and hence is already present for zero magnetic field, extending itself in the horizontal direction. The two transitions delimiting it are almost horizontal for small $\tilde{e}B$ and then move closer together until they meet around $\tilde{e}B\simeq0.085\;\mathrm{GeV^{2}}$ for both $H/G$ values, after which there is a unique transition connecting $B$ with $A$. The extension of this phase in the $\mu$ direction is only $10-15\: \mathrm{MeV}$ in range and it varies only slightly with $H/G$, but it is interesting to note that when charge neutrality effects are not taken into account, the $D$ phase does become wider when $H/G$ is increased. As we have seen, in this phase the densities are usually finite (except for the muonic one) with the LLL populated.

In the $A$ phase we see a series of near vertical first order transitions, whose origin lies in the quantization in LL's of the dispersion relations of quarks and leptons under the influence of magnetic fields, according to Eq.~(\ref{E1}). This behavior is in turn related to the already mentioned vA-dH effect, and is common to the populated phases of any NJL-type model with magnetic fields, where the medium contribution to the thermodynamical potential contains a sum over such LL's. In the basic NJL model, this effect is associated to all quark species, and the transitions occur when the maximum LL populated of a given species changes in one unit, giving rise to a weak jump in the order parameters. However, in the superconducting case with a rotated magnetic field under compact star conditions the behavior depends on the particle species, and only the $ub$ quarks and leptons exhibit ordinary vA-dH transitions as in the basic NJL model. On the other hand, $db$ quarks are not coupled to the field, so the form of the dispersion relation is that of the free quark, therefore not exhibiting the vA-dH effect. As for the paired $r$ and $g$ quarks, the corresponding medium sum in Eq.~(\ref{P_Delta}) is not cut off by a Heaviside function like for the $ub$ quark. This means that all of their LL's are populated, unless we find ourselves in the particular case $\Delta=0$ for which a Heaviside function is recovered. In the phase diagram, the $A$ phase is divided into sub-phases denoted as $A_i$, which in turn correspond to the phase where the $ub$ quark populates up to the $i$-th LL. If we traverse the phase diagram horizontally in the increasing $\tilde eB$ direction, the highest populated LL of a particular unit charged species decreases in one unit every time one of the corresponding transitions is crossed. In the $A_0$ phase, both $ub$ quarks and leptons are in the LLL. It should be noted that the maximum LL populated in these phase diagrams is smaller than the one obtained in: (a) the symmetric matter case, because $\mu_e$ reduces $\mu_{ub}$, and (b) the model without superconductivity, because $ \tilde{q}_{ub} > q_u,|q_d|$. Naturally, since $\mu_\mu\!=\!\mu_e$ and $m_\mu\!>\!m_e$, a lower value of $\mu_e$ is required to populate the electronic LL's. This is why there are more electronic transitions in the phase diagram, which occur for smaller values of $\tilde{e}B$ than the muonic ones. It is interesting to note that, in the case without superconductivity, the leptons exhibit an opposite behavior. There, since $\mu_e$ increases with the magnetic field, they become populated when the phase diagram is traversed in the increasing magnetic field direction~\cite{Allen:2015qxa}. 

Regarding the $g2SC$-type phases, we see that for $H/G=1$ both $A$ and $D$ phases are of the $2SC$-type. However, for $H/G=0.75$ there exists a region in the $A$ phase at low fields which is of the $g2SC$ type. The transition to the $2SC$ region is first order and occurs around $\tilde{e}B\simeq0.024\;\mathrm{GeV^{2}}$. Similarly, it is interesting to note that in the $D$ phase there is a small gapless region, bounded by first order transitions, which has already been discussed in Fig.~\ref{Fig2} for $\tilde{e}B=0.07\;\mathrm{GeV^{2}}$, and occurs when $\delta\mu > \Delta$. It is found in the intervals $0.062\;\mathrm{GeV^{2}}<\tilde{e}B<0.073\;\;\mathrm{GeV^{2}}$ and $342\;\mathrm{MeV}<\mu<344\;\mathrm{MeV}$. For the present value of $H/G$ this region is very small, but as we will discuss in the Appendix it is expected to grow rapidly when $H/G$ is decreased between $0.75$ and $0.7$.

The $P_{\delta\mu}$ term in Eq.~(\ref{P_deltamu}), which may be finite in the $g2SC$ phase, also contains a sum over LL's which is cut off by a Heaviside function. Therefore, this term will also give rise to transitions in the phase diagram, but their origin is different from that of the vA-dH ones. Subject to the condition that $\delta \mu\!>\!\Delta$, the densities of the $u$ and $d$ paired quarks will be equal if $k_{+}\!\le\!0$, where the $P_{\delta\mu}$ term will be zero. For $k_{+}\!>\!0$, on the other hand, $P_{\delta \mu}$ will be finite and the densities of the paired quarks will split. Furthermore, it can be noted that for a given $k_{+}$, the densities of the $k$-th Landau level will be different if $k\!<\!k_{+}$ and equal if $k\!>\!k_{+}$. Every time that $k_{+}$ increases in one unit, therefore, the densities of the corresponding Landau level split, signaled by the corresponding transition. However, the effect of these transitions is smeared out as $H/G$ is increased. Since they are numerous and become rather weak to affect the order parameters in a visible way, we have not included them in the phase diagram.

By comparing both diagrams for different values of $H/G$, it is easy to see how the phase diagram is modified when this ratio is swept between $0.75$ and $1$. Even though values of $H/G\gtrsim1$ are unlikely to be realized in QCD, the structure of the phase diagram would be maintained in such a case, with the difference that the first order transition between the $D$ and $A$ phases weakens and eventually turns into a crossover around $H/G\!\sim\!1.15$. For $H/G\!<\!0.75$, numerical difficulties arose when solving the system of equations, due to the fact that solutions with low $\Delta$ values are almost degenerate with the always present $\Delta\!=\!0$ solution. In order to shed some light on this issue, the $\tilde eB\!=\!0$ case is discussed in the Appendix. There, it is seen that the corresponding phase
structure can also present qualitative differences below $H/G\!=\!0.75$, such that the majority of the $D$ and $A$ phases can become gapless in a short range and eventually lose their superconducting behavior.

\subsection{Order parameters as a function of the magnetic field}

In order to provide some further understanding of the characteristics of the phases of the system under investigation we discuss here the behavior of the order parameters and densities as a function of the magnetic field. The results corresponding to Set~1 are shown in Fig.~\ref{Fig5}. Three different representative values of $\mu$ have been considered, chosen so as to include the different phases appearing in Fig.~\ref{Fig4}.

In the $B$ phase, $M$ always increases with the magnetic field according to the magnetic catalysis effect, which occurs principally in vacuum. For the lowest chemical potential in the diagram, and owing to the presence of the IMC well, an increase in $\tilde{e}B$ for constant $\mu$ first causes a transition from the $B$ phase to a restored phase, and a subsequent increase causes the system to return to the $B$ phase, which is always present for large enough fields. The latter effect takes place because, for large magnetic fields, the vacuum magnetic terms in Eqs.~(\ref{omegaone},\ref{P_Delta}) become dominant, favoring large masses. For intermediate chemical potentials
($\mu\!=\!342.5$ MeV for $H/G\!=\!0.75$ and $\mu\!=\!320$ MeV for $H/G\!=\!1$), the system is in the $D$ phase for low magnetic fields and in the $A_{0}$ phase for higher values, where the LLL of all species with charge $1$ is populated. In the $D$ phase, $M$ decreases slowly with $\tilde{e}B$, exhibiting the behavior of inverse magnetic catalysis. Once again, $\Delta$ and $\mu_{e}$ increase steadily, while $\mu_8$ tends to more negative values. It is interesting to note that it can change sign in the $A$ phase. For $\mu=360$ MeV, which corresponds to the $A$ phase for $\tilde{e}B\!<\!0.60$ GeV$^2$ approximately, the parameters exhibit oscillations and a series of peak-like discontinuities which correspond to the already discussed vA-dH transitions present in Fig.~\ref{Fig4}, and are similar to those discussed in~\cite{Allen:2015paa}.

We see that no oscillations are present in the $B$ phase, in contrast to previous studies which use smooth regularization functions~\cite{Fayazbakhsh:2010gc,Mandal:2012fq}. As mentioned before, the MFIR scheme removes these strong unphysical oscillations, also assuring that all of the oscillations present in the $A$ phase are real vA-dH transitions. Moreover, in this scheme the $B$ phase is always recovered for high enough magnetic fields, and the presence of an intermediate mixed phase whose existence depends on the set of parameters was revealed. 

The aforementioned vA-dH oscillations in the $A$ phase are related to the densities of charge 1 species as can be seen explicitly in Fig.~\ref{Fig6}, where we plot the densities of all species as a function of the magnetic field. From Eq.~(\ref{n_ub}), we see that they originate from the competition between the $\tilde eB$ factor in the expression for the density, which corresponds to the momentum degeneracy in the direction perpendicular to the field and increases with $\tilde{e}B$, and the phase space available in the $z$ direction for each LL, which decreases as a function of $\tilde{e}B$. The product of these terms results in density oscillations when $k>0$, which translate
into opposite oscillations for the dressed mass due to Pauli blocking effect. For $k=0$, $n_{ub}$ is almost proportional to the perpendicular momentum degeneracy, $|qB|$~\cite{Chakrabarty:1996te}, and increases steadily with the magnetic field growing faster than the densities of the paired quarks, while $n_{db}$ is explicitly independent of $\tilde{e}B$ (it depends slowly on $\tilde{e}B$ through $M$). To compensate this blue color excess, $\mu_8$ tends toward positive values for high magnetic fields, as seen in Fig.~\ref{Fig5} and also in Fig.~\ref{Fig2} for $\tilde{e}B=0.2\:\mathrm{GeV}^2$. This explains the decrease of $n_{db}$ with $\tilde{e}B$. Regarding the $g2SC$ mode we can check once again that the densities of the paired quarks get separated in this region, as shown in the inset of the intermediate left panel and in the lower left panel for small enough $\tilde{e}B$.

\section{Summary and Conclusions} \label{SecIV}

In the present work we explored the properties of magnetized cold color superconducting quark matter under compact star conditions within the framework of a two-flavor NJL-type model using the so-called ``magnetic field independent regularization'' (MFIR). Such regularization scheme was originally described in~\cite{Menezes:2008qt} and extended to the case with color pairing interactions in~\cite{Allen:2015paa}. The chiral and diquark condensates were obtained numerically iterating the coupled and self-consistent gap equations, under the constraints of color and charge neutrality as well as $\beta$-equilibrium. We considered two parameter sets that adjust to acceptable values of the dressed masses, and that were already known to generate qualitatively different phase diagrams in the symmetric matter case~\cite{Allen:2015paa}. Moreover, two representative values of the coupling constant ratio $H/G$ were considered. We presented the corresponding phase diagrams, where we found chirally broken ($B$) and (almost) restored ($A$) phases for both sets of parameters, connected through a first order transition. In particular, for Set~1 an intermediate mixed phase ($D$) was also found, where both condensates are large. It is bounded from below by a second order transition, from above by a first order one and disappears around $\tilde{e}B\simeq0.085\;\mathrm{GeV^{2}}$. We showed that both $A$ and $D$ phases are composed of $g2SC$ and/or $2SC$ regions, depending on the values of $H/G$, $\mu$ and $\tilde{e}B$. We described the behavior of the parameters for each phase as functions of $\tilde{e}B$ and $\mu$, and found that in the superconducting phases $\Delta$ can reach values of the order of 100 MeV for $H/G=0.75$ and 200 MeV for $H/G=1$, while $\mu_e<200\:\mathrm{MeV}$ and $|\mu_8|<70\:\mathrm{MeV}$. We also studied the density of the various particles present, and checked that in the $B$ phase all densities are null, in $A$ they are usually finite and in $D$ one finds quark and electron population even though $\mu<M$, while the muon density tends to be null. In addition, we verified that in the $2SC$ region the densities of the paired quark are equal, and usually get separated in the $g2SC$ one. The nature of the vA-dH transitions was also discussed. We explained that in the rotated base these exist only for the unit charged species, $ub$ quarks and leptons, and that even though the $|\tilde{q}|=1/2$ quarks couple to the magnetic field and present first order transitions in the $g2SC$ phase, their coupling to the superconducting gap smears these transitions out, becoming inappreciable for the chosen values of $H/G$. Furthermore, we found magnetic catalysis effects in vacuum and inverse magnetic catalysis (IMC) effects in the superconducting phases. As for other ratios of $H/G$, we commented that for $H/G\gtrsim1$ the structure of the phase diagram is maintained, with the difference that the phase transition between the $D$ and $A$ phases weakens and eventually turns into a crossover around $H/G\sim1.15$. Since numerical difficulties arose when solving the system of equations for $H/G<0.75$, we developed predictions for this regime based on the study of the $\tilde eB=0$ case. There we observed that, depending on the ratio $H/G$, both $A$ and $D$ phases could be in three different modes: (a) $2SC$ for $H/G\gtrsim 0.75$, (b) $g2SC$ for $0.65\lesssim H/G \lesssim 0.75$ and (c) $\Delta=0$ for $H/G \lesssim 0.65$. Although the set of parameters is not exactly the same, this result is in concordance with~\cite{Huang:2003xd}. From here we concluded that when $H/G$ is decreased under 0.75 in the presence of a magnetic field, the $A$ and $D$ phases are expected to become gapless and eventually have null superconducting gap. We recover then the $C$ phase present in~\cite{Allen:2015paa} with the difference that, when neutrality conditions are imposed, $\Delta$ is not small but null. The effect is more noticeable in the $A$ phase, where $\Delta$ was of the order of 40 MeV in the non-neutrality case.

We conclude that, as expected, several effects take place on the behavior of the cold magnetized quark matter when introducing compact star conditions. Their presence induces the existence of $g2SC$ and $2SC$ modes, and reduces the maximum Landau level reached in the vA-dH transitions. In general, charge neutrality constraints tend to reduce the superconducting effect, increasing the value of the critical chemical potentials and attenuating the magnetic catalysis effect, diminishing the depth of the IMC well. As a consequence, the phase diagram is moved upwards favoring different phases. In particular, for $H/G<0.65$ it is expected that the superconducting gap vanishes in the $C$ and $A$ phases. 

A possible next step in these studies would be the inclusion of vector interactions in the model. The recent observation of neutron stars with masses of approximately $2 M_\odot$~\cite{Demorest:2010bx,Antoniadis:2013pzd} places a strong limit on the equation of state of cold and dense matter. Studies performed including vector interactions typically lead to an increase of the calculated stellar masses~\cite{Klahn:2013kga,Menezes:2014aka,Pereira:2016dfg}, and it has been suggested that without these, the two-solar mass constraint cannot be satisfied. Therefore it could be interesting to study the effect of vector interactions on the phase diagram of magnetized color superconducting matter under compact star conditions. 

\section*{ACKNOWLEDGMENTS}

This work has been funded in part by CONICET (Argentina) under Grant No. PIP 578 and by ANPCyT (Argentina) under Grant No. PICT-2014-0492.

\section*{Appendix: H/G DEPENDENCE OF THE PHASE STRUCTURE IN THE ABSENCE OF MAGNETIC FIELD} \label{Appendix}

Although within the present type of models the impact of compact star conditions on the behavior of dense quark matter in the absence of magnetic fields was already analyzed almost 15 years ago~\cite{Huang:2003xd,Huang:2002zd}, we find it useful to review some issues concerning the parameter dependence of the results. In fact, most of the existing literature makes use of the particular choice $H/G=0.75$. Given the uncertainties associated with the determination of that quantity we find it convenient to present in this Appendix some results concerning the dependence of the model predictions as the diquark interaction strength is swept from zero to a realistic value.
As discussed in the main text, this provides some help to infer what might happen in some cases as an external
magnetic field is turned on. We will concentrate on results within Set~1. The thermodynamic potential is given in this case by
\begin{eqnarray}
\Omega_{MFA} &=&
\frac{(M-m_{0})^{2}}{4G}+\frac{\Delta^{2}}{4H}-\sum_{l=e,\mu}P_{MED}(\mu_l,m_l) -2P_{VAC}^{reg}(M)
\nonumber \\ &-&
\sum_{f=u,d}P_{MED}(\mu_{fb},M)-P_{\Delta}^{reg}(M,\Delta,\bar
\mu)-P_{\delta \mu}(M,\Delta,\bar \mu, \delta \mu),
\end{eqnarray}
where
\begin{eqnarray}
P_{\Delta}^{reg}(M,\Delta,\bar \mu) &=& \frac{2}{\pi^{2}}\int_{0}^{\Lambda}dp\ p^{2} \left(E_{\Delta}^{+}+E_{\Delta}^{-}\right), \\
P_{\delta\mu}(M,\Delta,\bar \mu, \delta \mu)&=&\frac{1}{\pi^2} \Theta(\delta \mu \!-\! \Delta) \sum_{s=\pm}s\:\Theta(\bar \mu^s \!-\! M) \int_0^{p^s} dp\ p^{2}\ (\delta \mu\!-\! E^-_\Delta).
\end{eqnarray}
Here, $\bar \mu^\pm \!=\! \bar \mu \pm \sqrt{\delta \mu^2 - \Delta^2} $, $p^\pm \!=\! \sqrt{(\bar{\mu}^\pm)^2 - M^2}$ and $E_{\Delta}^{\pm}=\sqrt{(\sqrt{p^{2}+M^{2}}\pm\bar{\mu})^{2}+\Delta^{2}}$, while $P_{VAC}^{reg}$ and $P_{MED}$ were defined in Eqs.~(\ref{P_vac}) and (\ref{P_med}) respectively.

The phase diagram in the $\mu-H/G$ plane is displayed in Fig.~\ref{Fig7}. As we can see, it is roughly divided into three regions in terms of the degree of symmetry breaking: the vacuum $B$ phase where chiral symmetry is broken, the $A$ phase where symmetry is restored, and a region of intermediate chemical potentials where restoration is partial. Here we see that in addition to the already found $D$ phase, a new intermediate phase appears, denoted as $C$ in concordance with previous studies. Their difference lies in the fact that, for finite magnetic field, the $C$ phase is separated from the $B$ one by a first order transition. The superconducting properties of these phases will naturally depend on the value of $H/G$. To the left of the diagram, in both $C$ and $A$ phases, the gap is zero, in agreement with what was discussed in Ref.~\cite{Huang:2003xd}. Actually, it can be shown that $\Delta=0$ is the only solution to the system of equations. It is interesting to note that, in the symmetric matter case, the gap is finite for arbitrarily small values of $H/G$~\cite{Shovkovy:2004me}. When a magnetic field is added this remains to be so, as seen in Ref.~\cite{Allen:2015paa} where $\Delta$ is rather small until around $H/G=0.5$. One can therefore expect that the $\Delta=0$ behavior present under compact star conditions is also mantained when an external magnetic field is imposed.

When $H/G$ is increased beyond 0.65, the system enters a superconducting phase in the $g2SC$ mode, which occurs for values in the intermediate range $0.65\lesssim H/G \lesssim 0.75$. Since $\Delta$ exhibits a very weak dependence with $\mu$ in the gapless mode of the $A$ phase, the transition is almost vertical, occurring at $H/G=0.64$. For intermediate chemical potentials, the transition from $C$ to $D$ occurs in the range $H/G\simeq0.64-0.7$, where the line is slightly tilted upwards and to the right. This tilting is reasonable since for lower values of $H/G$ it is expected that a higher chemical potential should be necessary to induce the gap. It is also noted that these transitions are second order. The $g2SC-2SC$ transition occurs for $H/G=0.77$ in the $A$ phase and for $H/G \simeq 0.72-0.75$ in the $D$ phase, where both transitions are second order again. Above this value, there are no qualitative differences in the diagram, except for the transition to the $A$ phase which turns from first order to crossover for $H/G>1.05$.

\begin{figure}
\centering{}\includegraphics[width=1\textwidth]{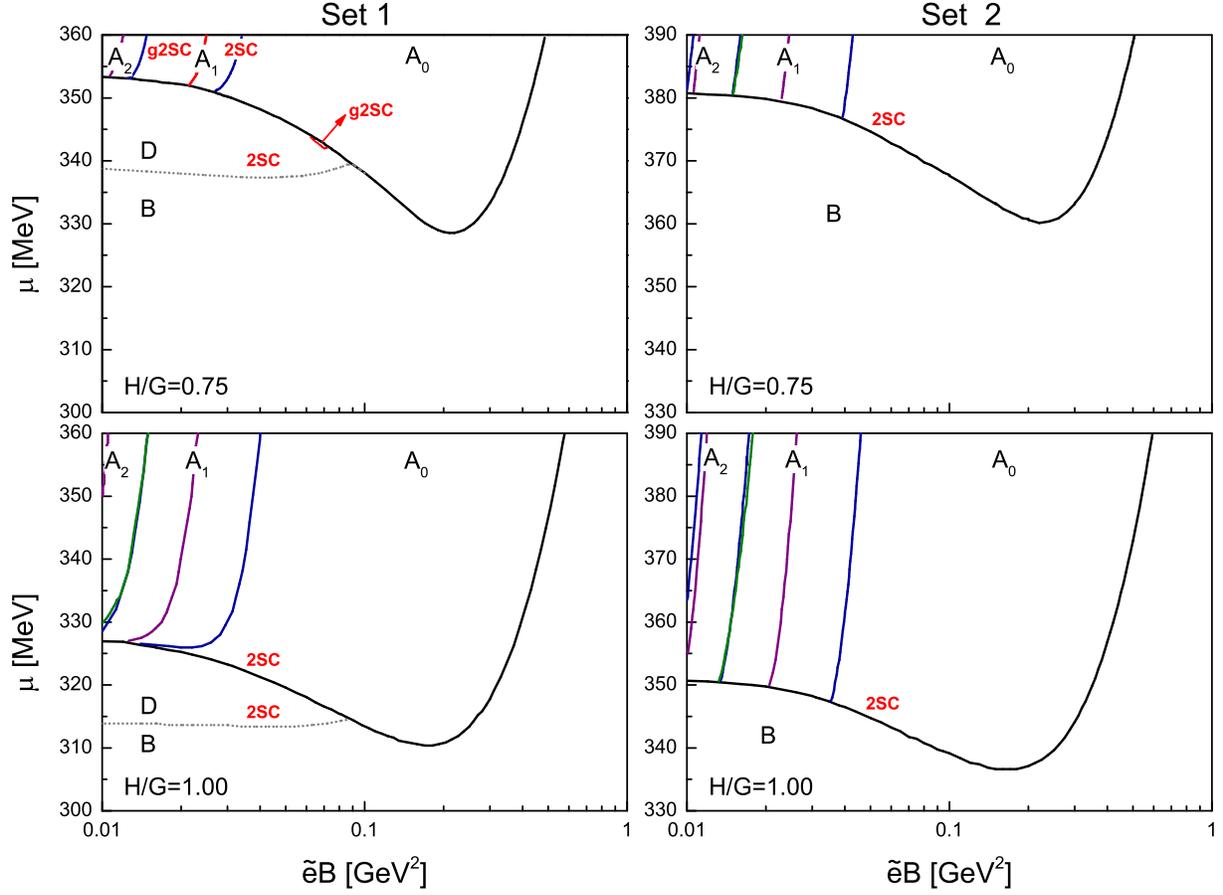}
\caption{(Color online) Phase diagrams for the two values of $H/G$ considered, for both Set~1 (left panels) and Set~2 (right panels). Full lines correspond to first order transitions and dotted lines to second order transitions. Black lines represent phase transitions. VA-dH transitions for unit charged species are colored as: blue for $ub$ quarks, violet for electrons and green for muons. Red lines indicate $g2SC$-$2SC$ transitions.  Note that the sets are plotted in different intervals of the $\mu$ axis, even though the scale is the same.}
\label{Fig4}
\end{figure}

\begin{figure}
\centering{}\includegraphics[height=0.9\textheight]{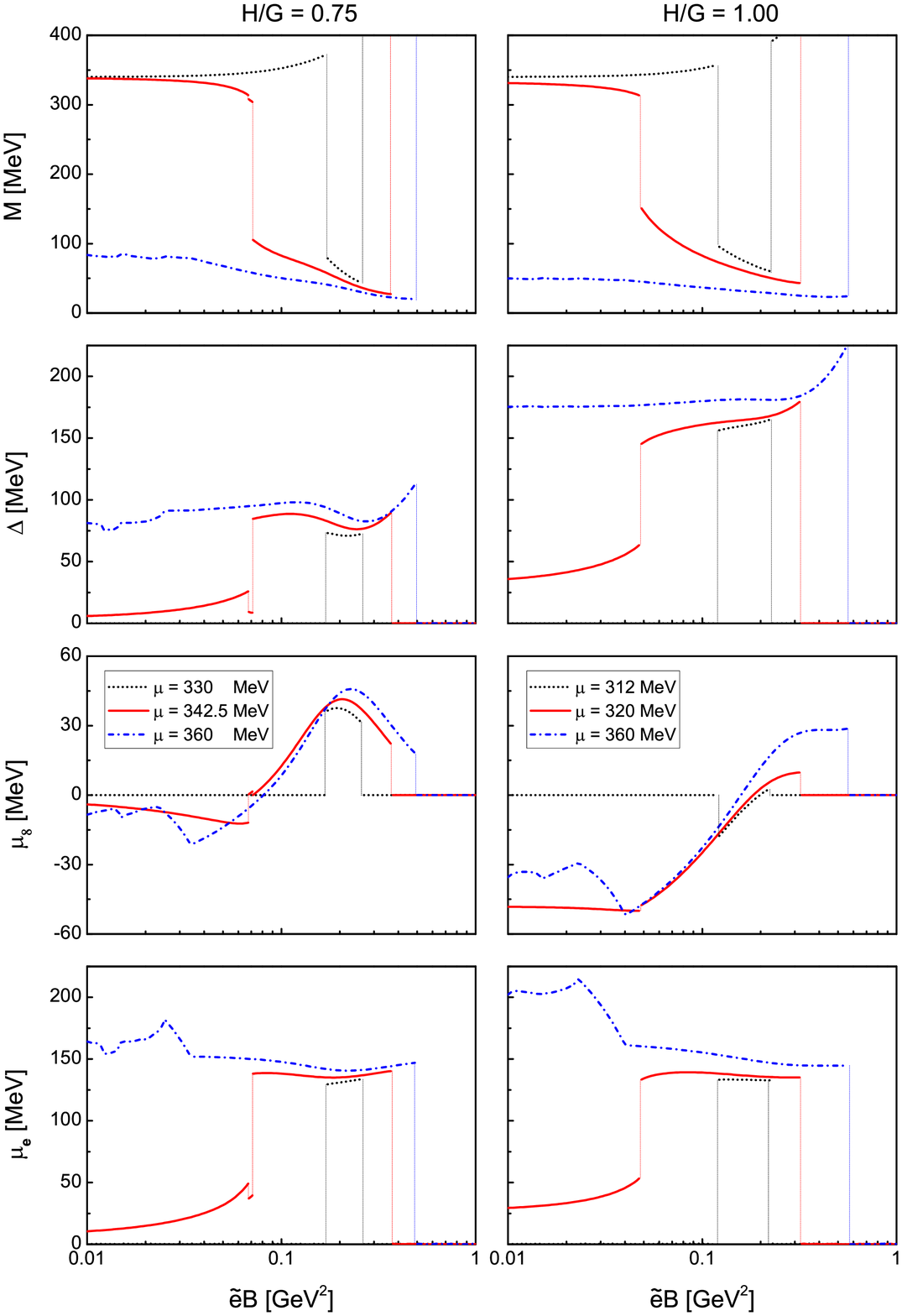}
\caption{(Color online) $M$, $\Delta$, $\mu_{8}$ and $\mu_{e}$ vs $\tilde{e}B$ for the two values of $H/G$ considered and three representative values of $\mu$, Set~1.}
\label{Fig5}
\end{figure}

\begin{figure}
\centering{}\includegraphics[height=0.9\textheight]{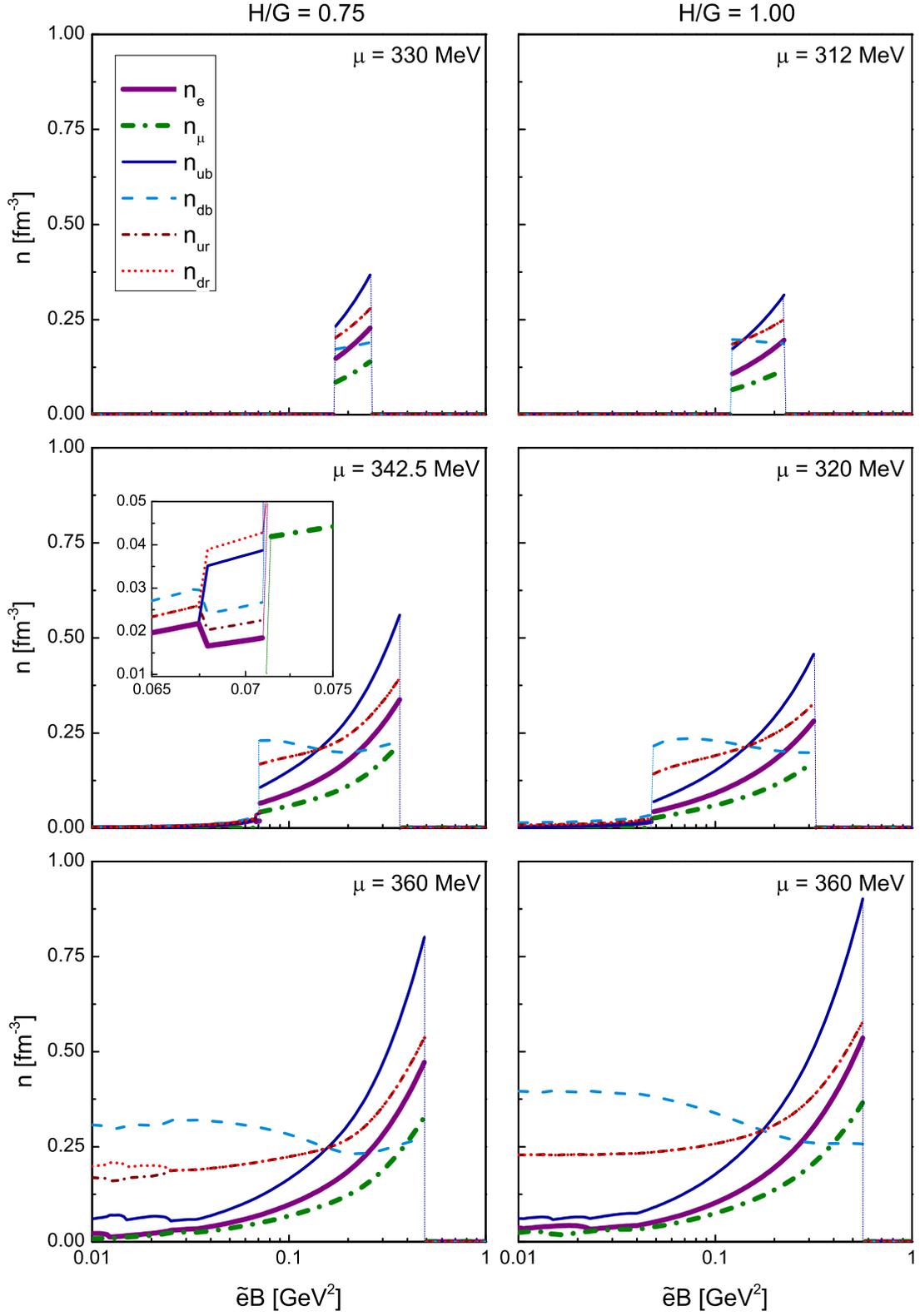}
\caption{(Color online) Quarks and leptons densities vs $\tilde{e}B$ for the two values of $H/G$ considered and three representative values of $\mu$ differents for each value of $H/G$, Set~1.}
\label{Fig6}
\end{figure}

\begin{figure}
\centering{}\includegraphics[width=0.8\textwidth]{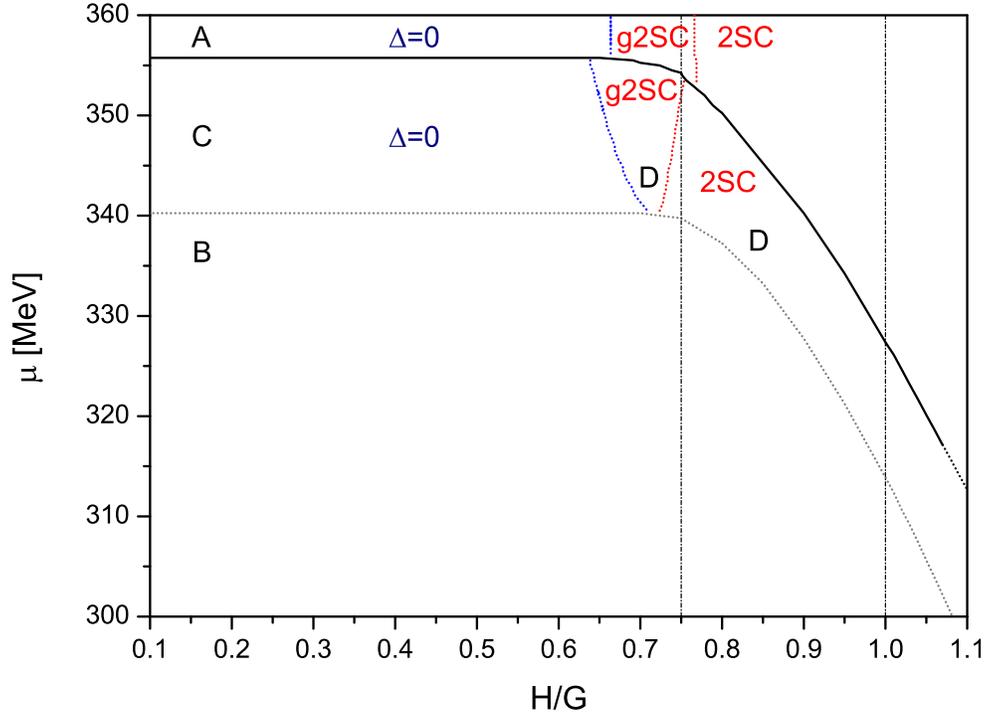}
\caption{(Color online) Phase diagram of $\mu$ vs $H/G$ for $\tilde{e}B=0$, Set~1. Full black lines correspond to first order transitions and dotted lines to second order transitions.}
\label{Fig7}
\end{figure}

It is interesting to observe from this phase diagram that $H/G=0.75$ corresponds to a very particular case in which the $D$ phase is found in the $2SC$ mode, while the $A$ phase is gapless. For this particular ratio, the model is therefore very sensitive to small changes of the parameters, a property that is expected to extend for the finite magnetic field case. As we saw in the corresponding phase diagram of Fig.~\ref{Fig4}, for $H/G=0.75$ and Set~1 there is a small gapless region inside the $D$ phase, in the range $0.062\:\mathrm{GeV^{2}}<\tilde{e}B<0.073\:\mathrm{GeV^{2}}$. This suggests that if $H/G$ is decreased from $0.75$, this region would grow horizontally toward the lower magnetic field zone (occupying first the upper part of the $D$ region), until the whole $D$ phase becomes gapless for $H/G=0.72$. Conversely, this would mean that if we start slightly to the right of the $g2SC-2SC$ transition, in the $2SC$ region, and increase the magnetic field from zero, the transition would displace toward the right, so that the system eventually returns to the gapless mode. In the $A$ phase, the behavior of the $g2SC-2SC$ transition is opposite. For zero magnetic field, the system is in the gapless mode for $H/G=0.75$, while in the phase diagram in the $\mu-\tilde eB$ plane there is a transition to the $2SC$ region around $\tilde eB\simeq 0.024\:\mathrm{GeV^2}$. Therefore, this transition is expected to move to lower magnetic field values when $H/G$ is increased, until it reaches the vertical axis for $H/G=0.77$. Once again, conversely, the transition line displayed in the $\mu-H/G$ plane would be displaced to the left when the magnetic field is increased, traversing the vertical line $H/G=0.75$ for $\tilde{e}B\simeq 0.024\:\mathrm{GeV^{2}}$. \\

\pagebreak

\end{document}